\documentclass[11pt]{article}
\usepackage{amsmath}
\usepackage{amssymb}
\usepackage{a4wide}
\numberwithin{equation}{section}
\title{\sc The Poisson-Sigma Model\\ A Non-linear Gauge Theory \footnote{Talk given at the {\it 3rd 
International Conference on Geometry, Integrability and Quantization} in Varna, Bulgaria, June 14-23 2001}}
\author{{\sc Thomas Schwarzweller}\footnote{\tt thomas@doom.physik.uni-dortmund.de}\\Fachbereich 
Physik\\Universit\"at Dortmund}

\begin{document}
%
%%%%%%%%%%%%%%%%%%%%%%%%%%%%%%%%%%%%%%%%%%%%%%%%%%%%%%%%%%%%%%%%%%%%%%%%%%%%%%%%%%%%%%%%%
%
\newcommand{\sss}{\scriptscriptstyle}
\newcommand{\mc}{\mathcal}
\newcommand{\beq}{\begin{equation}}
\newcommand{\eeq}{\end{equation}}
\newcommand{\rd}{\mathrm{d}} % Roman d for differential
\newcommand{\re}{\mathrm{e}} % Roman e for exponential
\newcommand{\ri}{\mathrm{i}} % Roman i for imaginary
%
%%%%%%%%%%%%%%%%%%%%%%%%%%%%%%%%%%%%%%%%%%%%%%%%%%%%%%%%%%%%%%%%%%%%%%%%%%%%%%%%%%%%%%%%%
%
\maketitle
\begin{abstract}
I investigate the {\sc Poisson}-sigma model on the classical and quantum level. First I show how the 
interaction can be obtained by a deformation of the classical master equation of an Abelian BF theory in 
two dimensions. 
On the classical level this model includes various known two-dimensional field theories, in particular 
the {\sc Yang-Mills} theory. On the quantum level the perturbation expansion of the path integral in the 
covariant gauge yields the {\sc Kontsevich} deformation formula. Finally I perform the calculation of the 
path integral in a general gauge, and demonstrate how the derived partition function reduces in the special 
case of a linear {\sc Poisson} structure to the familiar form of 2d {\sc Yang-Mills} theory.
\end{abstract}
%
%
%%%%%%%%%%%%%%%%%%%%%%%%%%%%%%%%%%%%%%%%%%%%%%%%%%%%%%%%%%%%%%%%%%%%%%%%%%%%%%%%%%%%%%%%%
%
\section{Introduction}
The class of non-linear gauge theories introduced by {\sc Ikeda} in \cite{I} is  based on a polynomial 
extension of the underlying {\sc Lie} algebra, for instance a finite W-algebra or a {\sc Poisson} algebra. 
Due to this non-linearity these models involve in the language of gauge theories an open gauge algebra, 
i.e. the algebra closes only on-shell. In such cases neither the {\sc Faddeev-Popov} quantization nor the 
BRST procedure leads to an appropriate application of a path integral quantization since both needs a 
well-defined cohomology to construct physical variables but the corresponding BRST operator is only 
nilpotent modulo the equations of motion. The proper method that works in these cases is the 
{\sc Batalin-Vilkovisky} procedure, for a detailed description see \cite{GPS}. This formalism has a 
beautiful geometric interpretation that enables one to receive the extended action used in the path 
integral from fundamental geometric ingredients, a nilpotent vector field and a symplectic structure on 
appropriate super manifolds \cite{AKSZ}. A famous representative of a non-linear gauge theory is the 
{\sc Poisson}-sigma model \cite{SS} which obviously is a sigma model with {\sc Poisson} structure 
on the target manifold. The interest in the last years is essentially based on its connection to the  
{\sc Kontsevich} formula for deformation quantization of {\sc Poisson} manifolds \cite{K} 
discovered by {\sc Cattaneo} and {\sc Felder} \cite{CF1}. Another important aspect is that the classical model associates 
to certain {\sc Poisson} structures on a finite-dimensional manifold two-dimensional field theories.
For a linear {\sc Poisson} structure which leads to the two-dimensional {\sc Yang-Mills} theory it was 
possible to show this connection also on the quantum level, i.e. the partition function of both 
theories corresponds to each other \cite{HS1}.

Also based on the {\sc Batalin-Vilkovisky} procedure, strictly speaking on the classical master equation, 
is a method to generate a physical consistent interaction between the gauge fields of a free gauge theory 
\cite{BH}. This problem then turns into a deformation of the solution of the master equation which is 
possible since it contains all the information about the considered gauge structure. 

This article is structured as follows. In section 2 the classical theory of the {\sc Poisson}-sigma model 
will be presented. That includes a short review of the general method to construct an interaction and the 
derivation of the model by means of a deformation of a free BF theory. In section 3 then the quantization 
will be performed, in particular the {\sc Feynman} perturbation series \cite{CF1} and the non perturbative 
calculation of the partition function \cite{HS1}.
%
%
%%%%%%%%%%%%%%%%%%%%%%%%%%%%%%%%%%%%%%%%%%%%%%%%%%%%%%%%%%%%%%%%%%%%%%%%%%%%%%%%%%%%%%%%%
%
\section{The Classical Theory}
%
%%%%%%%%%%%%%%%%%%%%%%%%%%%%%%%%%%%%%%%%%%%%%%%%%%%%%%%%%%%%%%%%%%%%%%%%%%%%%%%%%%%%%%%%%
%
\subsection{Consistent Interaction via Deformation}
In this section the cohomological approach to the problem of generating consisting interactions is reviewed 
\cite{BH}. It essentially is based on the antifield formalism in the sense that a deformation of the 
solution of the master equation leads to an action functional containing a consistent interaction term.
 
First of all I present a very brief review of the antifield formalism. Indeed, here are 
just some properties which will be needed in the forthcoming sections. The starting 
point is an action $S_0[\phi]$ with gauge symmetries $\delta_\epsilon\phi^i=
\mc{R}^i_\alpha\epsilon^\alpha$. By introducing ghosts and antifields one can construct a solution 
$S[\phi^A,\phi^\ast_A]$ of the master equation
\beq
(S,S)=\frac{\delta S}{\delta\phi^A}\frac{\delta S}{\delta\phi^\ast_A}-\frac{\delta S}
{\delta\phi^\ast_A}\frac{\delta S}{\delta\phi^A}=0,\;\;\;\;{\rm  with}\;\;
S=S_0+\phi^\ast_i\mc{R}^i_\alpha C^\alpha +\cdots\,,
\end{equation}
where $\phi^A=(\phi^i,C^\alpha,...)$ denotes the sets of the original fields, the ghosts and so on, while 
$\phi^\ast_A$ stands for the corresponding antifields. In the solution of the master equation the whole 
information about the gauge structure is encoded. The algebra of the fields and antifields together with 
the BRST-differential generated by $S$ through the antibracket
\beq
s(\cdot)=(\cdot,S)
\end{equation}
yields a complex. The corresponding BRST cohomology is denoted by $H^\star=\sum H^p(s)$. One can define a 
map in the cohomology induced by the antibracket, the {\it antibracket map}
\beq
(\cdot,\cdot ):H^p(s)\times H^q(s)\longrightarrow H^{(p+q+1)}(s)\;,
\end{equation}
\beq
([A],[B])\mapsto [(A,B)]\,.
\end{equation}
The important result found by {\sc G.Barnich} and {\sc M.Henneaux} is the fact that the antibracket map is 
trivial in the sense the antibracket of 2 BRST-closed functionals is BRST-exact, for a proof consult 
\cite{BH}.  Due to the triviality of the antibracket map one can define higher order maps in the 
cohomology, however it turns out they are trivial in a similar way.

Now consider a {\it free} gauge theory with a {\it free} symmetry given by 
\[{\rm  Free\; Theory}=\left\{ \begin{array}{l}
{\rm  Free\; action:\;}{\stackrel{\sss (0)}{S}_{ 0}[\phi^i]}\\
{\rm Gauge\; Symmetry:\;}{\delta_\epsilon\phi^i=\stackrel{\sss(0)}{{\mathcal R}^i}_\alpha\epsilon^\alpha}\\
{\rm Noether\; Theorem:\;}{\frac{\delta\stackrel{\sss (0)}{S}_{0}[\phi^i]}{\delta\phi^i}\stackrel{\sss (0)}
{{\mathcal R}^i}_\alpha=0}\;.
\end{array}\right.\]
The aim is now to introduce couplings between the fields $\phi^i$ which fulfill the crucial physical 
requirement of preserving the number of gauge symmetries, those couplings will be called consistent. 
It means one has to perturb the action and the symmetries
\beq
\stackrel{\sss(0)}{S}_{ 0}\longrightarrow S_{ 0}=\stackrel{\sss(0)}{S}_{ 0}+
g\stackrel{\sss(1)}{S}_{ 0}+g^2\stackrel{\sss(2)}{S}_{ 0}+...\;\;,
\end{equation}
\beq \stackrel{\sss(0)}{{\mathcal R}^i}_\alpha\longrightarrow {\mathcal R}^i_\alpha=
\stackrel{\sss(0)}{{\mathcal R}^i}_\alpha+g\stackrel{\sss(1)}{{\mathcal R}^i}_\alpha+
g^2\stackrel{\sss(2)}{{\mathcal R}^i}_\alpha+...\;\;,
\end{equation}
such that $\delta_\epsilon\phi^i={\mathcal R}^i_\alpha\epsilon^\alpha$ is a 
symmetry of $S_0$
\beq \frac{\delta(\stackrel{\sss (0)}{S}_{ 0}+g\stackrel{\sss(1)}{S}_{ 0}+
g^2\stackrel{\sss(2)}{S}_{ 0}+...)}{\delta\phi^i}
(\stackrel{\sss(0)}{{\mathcal R}^i}_\alpha+g\stackrel{\sss(1)}{{\mathcal R}^i}_\alpha+
g^2\stackrel{\sss(2)}{{\mathcal R}^i}_\alpha+...)=0\,,
\end{equation}
which expresses the consistency. It is not an easy task to deform simultaneously the action and the 
symmetry to get a consistent interaction.

This problem can be reformulated as a deformation problem of the solution of the master equation. Basically 
this procedure is based on the fact that the master equation contains all the information about the gauge 
structure
\beq
(\stackrel{\sss(0)}{S}_{BV},\stackrel{\sss(0)}{S}_{BV})=0\longrightarrow (S_{BV},S_{BV})=0\;,
\end{equation}
\beq
\stackrel{\sss(0)}{S}_{BV}\longrightarrow S_{ BV}=\stackrel{\sss(0)}{S}_{BV}
+g\stackrel{\sss(1)}{S}_{BV}+g^2{\stackrel{\sss(2)}{S}}_{BV}+...\;.
\end{equation}
The deformed master equation guarantees now the consistency of $S_0$ and ${\mathcal R}^i_\alpha$ and further on the 
original and the deformed gauge theory have the same spectrum of ghosts and antifields.

The advantage of this formulation is that one now can use the cohomological techniques of deformation 
theory. The deformed master equation can be analyzed order by order in the deformation parameter, the 
coupling constant. This expansion yields the following relations
\beq
(\stackrel{\sss(0)}{S}_{ BV},\stackrel{\sss(0)}{S}_{ BV})=0\label{exp1}\;,
\end{equation}
\beq 2(\stackrel{\sss(0)}{S}_{ BV},\stackrel{\sss(1)}{S}_{ BV})=0\label{exp2}\;,
\end{equation}
\beq2(\stackrel{\sss(0)}{S}_{ BV},\stackrel{\sss(2)}{S}_{ BV})+(\stackrel{\sss(1)}{S}_{ BV},
\stackrel{\sss(1)}{S}_{ BV})=0\label{exp3}\;,
\end{equation}
\[({\rm + higher\; orders})\;.\]
The first equation (\ref{exp1}) is fulfilled by assumption, it is exactly the master equation for the free 
gauge theory. Equation (\ref{exp2}) shows that $\stackrel{\sss(1)}{S}_{ BV}$ is forced to be a cocycle of 
the free BRST differential $\stackrel{\sss(0)}{s}$. Assume now $\stackrel{\sss(1)}{S}_{ BV}$  is a 
coboundary then the corresponding interaction belongs to a field redefinition which should not be 
considered and the deformation will be called trivial. Therefore $\stackrel{\sss(1)}{S}_{ BV}$ is an 
element of the zeroth cohomological space $H^0(\stackrel{\sss(0)}{s})$ which is isomorphic to the space of 
physical observables of the free theory. Because of the triviality of the antibracket map 
$(\stackrel{\sss(1)}{S}_{ BV},\stackrel{\sss(1)}{S}_{ BV})$ is BRST exact and one gets no obstructions 
for constructing the interaction from (\ref{exp3})  and $\stackrel{\sss(2)}{S}_{ BV}$ exist. This is also 
true for higher orders, so there are no obstructions for the interacting action at all.

Usually the original action is a local functional of a corresponding {\sc Lagrange} function, such that 
also the deformations need to be local functionals. Taking locality into account the analysis gets much 
more involved because the antibracket map is not trivial anymore, e.g. the antibracket of two local BRST 
cocycles need not necessarily to be the BRST variation of a local functional. Consider 
$\stackrel{\sss(k)}{S}=\int\stackrel{\sss(k)}{\mc{L}}$, where $\mc{L}$ is the Lagrangian, an n-form. 
The corresponding (local) antibracket is defined modulo an d-exact term, d being the exterior derivative. 
This yields for the deformation expansion of the Lagrangian
\beq
2\stackrel{\sss(0)}{s}\stackrel{\sss(1)}{\mc{L}}=\rd\stackrel{\sss(1)}{j}\;,
\end{equation}
\beq
\stackrel{\sss(0)}{s}\stackrel{\sss(2)}{\mc{L}}+\{\stackrel{\sss(1)}{\mc{L}},\stackrel{\sss(1)}{\mc{L}}\}
=\rd\stackrel{\sss(2)}{j}\,,
\end{equation}
\[({\rm + higher\; orders})\,,\]
where $\stackrel{\sss(k)}{j}$ is the symbol for the d-exact term.
$\stackrel{\sss(1)}{\mc{L}}$ is BRST closed modulo d, this means that the nontrivial deformations of the 
master equations belong to $H^0(\stackrel{\sss(0)}{s}\mid\rd)$. Because the corresponding local 
antibracket is no longer trivial, it possesses a lot of structure, one gets obstructions for the 
construction of the interaction term, some so-called consistency conditions. The construction of local 
consistent interaction is strongly constrained.
%
%%%%%%%%%%%%%%%%%%%%%%%%%%%%%%%%%%%%%%%%%%%%%%%%%%%%%%%%%%%%%%%%%%%%%%%%%%%%%%%%%%%%%%%
%
\subsection{Deformation of Abelian BF Theory}
As an application of the formalism consider now the nontrivial deformation of Abelian BF theory in two 
dimensions \cite{IZ}. The free action is given by
\beq
\stackrel{\sss (0)}{S}_0=\int\limits_{\Sigma_g}\; A_i\wedge \rd\phi^i\;,
\end{equation}
which is invariant under the gauge transformation
\beq
\delta\phi^i=0,\;\;\;\;\;\delta A_i=\rd\epsilon_i\;.
\end{equation}
The minimal solution of the classical master equation is 
\beq
\stackrel{\sss (0)}{S}=\int\limits_{\Sigma_g}[A_i\wedge \rd\phi^i+ A^{\ast i}\rd C_i]
\end{equation}
and the corresponding BRST-Symmetry is 
\beq
\stackrel{\sss (0)}{s}=\rd C_i\frac{\vec{\partial}}{\partial A_i}+\rd\phi^i
\frac{\vec{\partial}}{\partial A^{\ast i}}+\rd A_i\frac{\vec{\partial}}
{\partial \phi^\ast_i}-\rd A^{\ast i}\frac{\vec{\partial}}{\partial C^{\ast i}}\,.
\end{equation}
The first order deformation $\stackrel{\sss(1)}{\mathcal L}$ of the Lagrangian  associated to the 
minimal solution $\stackrel{\sss(0)}{S}$ should obey the following condition
\beq
\stackrel{\sss(0)}{s}\stackrel{\sss(1)}{\mathcal L}+\rd a_{[1]}=0\;.
\end{equation}
It defines an element of $H^0(\stackrel{\sss (0)}{s}\mid\rd)$, so that one 
gets a set of {\it descent equations}
\beq
\stackrel{\sss (0)}{s}a_{[1]}+\rd a_{[o]}=0\;,\;\;\;\stackrel{\sss (0)}{s}a_{[0]}=0\;.
\end{equation}
It is a simple calculation to get the solution for $\stackrel{\sss(1)}{\mathcal L}$
\begin{multline}
 \stackrel{\sss (1)}{\mathcal L}=-\frac{1}{4}\frac{\delta^2 f^{ij}[\phi]}
{\delta\phi^k\delta\phi^l}A^{\ast k}\wedge A^{\ast l}C_i C_j
+\frac{\delta f^{ij}[\phi]}{\delta\phi^k}C^{\ast k}C_i C_j\\
-\frac{\delta f^{ij}[\phi]}{\delta\phi^k}A^{\ast k}\wedge A_i C_j 
-f^{ij}[\phi]\phi^{\ast}_i C_j + \frac{1}{2}f^{ij}[\phi]A_i\wedge A_j\,.
\end{multline}
The $f^{ij}[\phi]$ are antisymmetric and to receive a consistent interaction they have to verify
\beq
\sum\limits_{cycl(ijk)} \frac{\delta f^{ij}[\phi]}{\delta \phi^l}f^{kl}[\phi]=0\label{cc}\;,\end{equation}
which is a generalized {\sc Jacobi} identity. Since this condition is fulfilled there are no obstructions 
in the construction and the second order deformation can be chosen to be zero.
This yields for the deformed solution of the master equation
\begin{equation}
\begin{split}
S_{BV}=\stackrel{\sss(0)}{S}+\stackrel{\sss(1)}{S}=& \int\limits_{\Sigma_g}
\bigr[ A_i\wedge\rd\phi^i-{ \frac{\delta f^{ij}[\phi]}{\delta\phi^k}
A^{\ast k}\wedge A_i C_j}+A^{\ast i}\wedge\rd C_i-{ f^{ij}[\phi]
\phi^{\ast}_i C_j}\\
& \phantom{\int\biggr[}+ \frac{1}{2}f^{ij}[\phi]A_i\wedge A_j
+\frac{\delta f^{ij}[\phi]}{\delta\phi^k}C^{\ast k}C_i C_j 
-\frac{1}{4}\frac{\delta^2 f^{ij}[\phi]}{\delta\phi^k\delta\phi^l}A^{\ast k}\wedge 
A^{\ast l}C_i C_j\bigr]\;.
\end{split}\label{exa}
\end{equation}
From this extended action one can read off the classical action including an interaction term quadratic in  
the gauge fields $A_i$
\beq S_{0}=\stackrel{\sss(0)}{S_0}+\stackrel{\sss(1)}{S_0}=\int\limits_{\Sigma_g}
\bigr[A_i\wedge\rd\phi^i+{ \frac{1}{2}f^{ij}[\phi]A_i\wedge A_j}\bigr]\;\end{equation}
and the deformed gauge symmetries are
\beq\delta_\epsilon \phi^i={ f^{ji}[\phi]\epsilon_j}\;,\end{equation}
\beq\delta_\epsilon A_i=\rd\epsilon_i -{  f^{kl}{}_{,i}[\phi]A_k\wedge A_l}\,.\end{equation}
Note, the gauge algebra is only closed on shell, which reflects the non-linearity of the gauge algebra.
%
%%%%%%%%%%%%%%%%%%%%%%%%%%%%%%%%%%%%%%%%%%%%%%%%%%%%%%%%%%%%%%%%%%%%%%%%%%%%%%%%%%%%%%%%%
%
\subsection{The Poisson-Sigma Model}
The {\sc Poisson}-sigma model is a two-dimensional gauge theory  based on a {\sc Poisson} algebra, 
i.e. the target manifold of the theory is a {\it Poisson manifold} $(N,P)$, that is a smooth manifold $N$ 
equipped with a {\sc Poisson} structure $P\in\Lambda^2TN$. In local coordinates $X^i$ on $N$ the 
{\sc Poisson} structure is given by
\beq
P=\frac{1}{2}P^{ij}(X)\partial_i\wedge\partial_j
\end{equation}
and $P^{ij}(X)$ has to satisfy the {\sc Schouten-Nijenhuis} constraint
\beq
P^{i[j}(X)P^{kl]}{}_{,i}(X)=0\;,
\end{equation}
where $[jkl]$ stands for a cyclic sum. That is exactly the consistency condition (\ref{cc}) for the 
structure functions of the deformed two dimensional BF theory. Thus the {\sc Poisson}-sigma model is a 
realization of the deformed Abelian BF theory.

It is possible to extend the action in such a way that the symmetries are unchanged. The {\sc Poisson}
structure induces a map $T^\star N\rightarrow TN$ which is not surjective (like in the symplectic case). 
However, due to the {\sc Jacobi} identity the image of this map forms an involutive distribution of vector 
fields. Further the associated characteristic distribution is completely integrable and the {\sc Poisson}
structure induces a symplectic structure $\Omega_L$ on the leaves L. It is a fact that a {\sc Poisson}
manifold is the disjoint union of its symplectic leaves. The {\it splitting theorem} of {\sc Weinstein} 
states that for regular {\sc Poisson} manifolds there exist so-called {\sc Casimir-Darboux} coordinates 
locally. The {\sc Poisson} manifolds under consideration in the following sections are isomorphic to 
$\mathbb{R}^n$. For $P$ degenerate there are non vanishing functions $f$ on $N$ whose {\sc Hamilton}ian 
vector fields $X_f\equiv f_{,i}P^{ij}\partial_j$ vanish, the {\sc Casimir} functions. Then 
$C^I(X)=const.=C^I(X_0)$ defines a level surface through $X_0$ whose connected components may be identified 
with the symplectic leaf $L$. This yields natural coordinates $\{X^I,X^\alpha\}$ on $(N,P)$, 
where $\{X^I\}$ is a whole set of {\sc Casimir} functions while $X^\alpha$ stands for the {\sc Darboux} 
coordinates on the leaf $L$ with $P^{IJ}=P^{I\alpha}=0$ and $P^{\alpha\beta}=(\Omega^{-1})^{\alpha\beta}$. 
These coordinates will be quite useful in calculating the partition function of the model.
The invariant nature of the {\sc Casimir} functions enables one to add it to the action but the topological 
invariance of the theory is broken due to the volume form $\mu$ of the world sheet $\Sigma_g$. The action 
and the symmetries are then
\beq
S[X,A]=\int\limits_{\Sigma_g} [A_i\wedge\rd X^i+\frac{1}{2}P^{ij}[X]A_i\wedge A_j+\mu C[X]]\;,
\end{equation}
\beq
\delta_\epsilon X^i=f^{ji}[X]\epsilon_j\;\;\;\;\;
\delta_\epsilon A_i=-\rd\epsilon_i + P^{kl}{}_{,i}[X]A_k\wedge A_l={\mathcal D}\epsilon_i\;,
\end{equation}
it is the action of the {\it Poisson-sigma model} introduced in \cite{SS}.

An interesting aspect of this model is that it associates to certain {\sc Poisson} structures on a 
finite-dimensional manifold  two-dimensional field theories. First consider the case in which the 
{\sc Poisson} structure $P$ gives rise to a symplectic 2-form on N, i.e. it has an inverse $\Omega$. Then 
it is possible to eliminate the gauge fields $A_i$ by means of the equations of motion and the resulting 
action is
\beq
S_{top}=\int\limits_{\Sigma_g}\Omega_{ij}\rd X^i\wedge\rd X^j\;,
\end{equation}
which evidently is the action of {\it Witten's topological sigma Model}.

Secondly choose a linear {\sc Poisson} structure $P^{ij}=c^{ij}_kX^k$ on $N=\mathbb{R}^3$. The coefficients 
$c^{ij}_k$ then define a {\sc Lie} algebra on the dual space which in turn can be identified with $N$. 
Now there are two different {\sc Casimir} functions. The trivial one $C=0$ leads back to the two-dimensional 
BF theory,  although now in the non Abelian case. On the other hand if one chooses now the quadratic 
{\sc Casimir} function $C=X^iX^i$ one ends with 
\beq
S_{YM}=\int\limits_{\Sigma_g} F^i\wedge\ast F_i\;,
\end{equation}
the {\it two-dimensional Yang-Mills theory}. The {\sc Poisson}-sigma model also covers the gauged 
{\sc Wess-Zumino-Witten} model and a theory of gravitation \cite{SS}.
%
%
%%%%%%%%%%%%%%%%%%%%%%%%%%%%%%%%%%%%%%%%%%%%%%%%%%%%%%%%%%%%%%%%%%%%%%%%%%%%%%%%%%%%%%%%%
%
%
\section{Quantization of the Poisson-sigma Model}
In this section the quantum theory of the {\sc Poisson}-sigma model is under consideration. For the {\sc 
Dirac} quantization I refer to \cite{SS}. 
%
%%%%%%%%%%%%%%%%%%%%%%%%%%%%%%%%%%%%%%%%%%%%%%%%%%%%%%%%%%%%%%%%%%%%%%%%%%%%%%%%%%%%%%%%%
%
\subsection{The Perturbative Series}
This first part is concerned with the perturbation expansion of the path integral of the 
{\it topological} {\sc Poisson}-sigma model which yields the {\sc Kontsevich} quantization formula \cite{CF1}. 
It turns out as the correlation function of two functions with support on the boundary of the 
two-dimensional disc, which is the topology of the world sheet
\beq
(f\star g)(x)=\int\limits_{X(\infty)=x}\;[D X][D A]\; f(X(1))\,g(X(0))
\exp(\frac{\ri}{\hbar}S[X,A])\;,
\end{equation}
where $(0,1,\infty)$ are 3 arbitrary, cyclic ordered points on the boundary of the disc. The boundary 
conditions for all fields are discussed in detail in \cite{CF1}.
 
In \cite{CF2} it was shown that the {\sc Batalin-Vilkovisky} action can be obtained by the so-called AKSZ-
formalism \cite{AKSZ}, a geometrical construction for the extended action without a classical action 
as a starting point. The crucial resulting point is one gets a set of generalized maps, the superfields
\beq
\tilde{X}^i=X^i+\theta^\mu A^{\ast i}_\mu-\frac{1}{2}\theta^\mu\theta^\nu C^{\ast i}_{\mu\nu}\;,
\end{equation}
\beq
\tilde{A}_i=C_i+\theta^\mu A_i+\frac{1}{2}\theta^\mu\theta^\nu X^{\ast}_{i\mu\nu}\;,
\end{equation}
where $\theta$ are {\it fermionic} coordinates of a corresponding super manifold. The action in this 
extended {\it super space} can be formulated as 
\beq
 S_{ BV}=\int\limits_{D^2}\int\rd^2\theta \bigl[ \tilde{A}_i{\mathcal D}\tilde{X}^i+\frac{1}{2}
P^{ij}[\tilde{X}]\tilde{A}_i \tilde{A}_j\bigr]\;.
\end{equation}
The integration over the fermionic coordinates is in the {\sc Berenzin} sense and $\mc{D}=\theta^\mu
\partial/\partial u^\mu$. Note that in this formalism the {\it formal} form of the action is similar to the 
original classical one of the {\sc Poisson}-sigma model but now the ghosts and the antifields are 
included in the superfields, in fact after integrating out the fermionic coordinates $\theta$ it is the 
extended action of the {\sc Batalin-Vilkovisky} formalism (\ref{exa}).

Sill in this approach one needs a gauge fixing procedure since the extended action possesses gauge 
invariances. In the {\sc Batalin-Vilkoviski} formalism the main idea is to introduce a {\it gauge fermion}, 
a functional of the fields only and with ghost number -1. The integral then is taken over the {\it 
Lagrangian} sub manifold $\Sigma_\Psi$ defined by $\Phi^\ast_A=\partial\Psi/\partial\Phi^A$. The problem is 
to find a functional $\Psi$ making the integral well-defined. To construct a gauge fermion of ghost number 
-1 one must introduce additional fields. The simplest choice is a {\it trivial} pair $\bar{C}_i,\bar{\pi}_i$, 
being  {\sc Lagrange} multiplier fields $\bar{\pi}_i$ having ghost number 0 and the {\sc Faddeev-Popov} 
antighosts $\bar{C}_i$ with ghost number -1 plus the associated antifields. The corresponding action of 
these fields is $-\int\bar{\pi}_i\bar{C}^{\ast i}$.
For calculating the perturbation expansion the natural choice of a gauge is the {\sc Feynman} gauge 
$\rd\star A_i=0$ resulting from the gauge fermion $\Psi=-\int\;\rd\bar{C}^i\star A_i$
Using this gauge fermion the {\it gauged fixed} action takes the form
\begin{multline}
S_{ gf}= \int\limits_{D^2} \bigl[A_i\wedge\rd X^i+\frac{1}{2}P^{ij}[X]A_i
\wedge A_j-\star\rd\bar{C}^i\wedge(\rd C_i+P^{kl}{}_{,i}[X] A_kC_l)\\
-\frac{1}{4}\star\rd\bar{C}^i\wedge\star\rd\bar{C}^j 
P^{kl}{}_{,ij}[X] C_kC_l-\pi^i\rd\star A_i \bigr]\;.
\end{multline}
The perturbation expansion is calculated by 
\beq
\int\limits_{\Sigma_\psi}[D\Phi^\ast_A][D\Phi^A]\exp(\frac{\ri}{\hbar} S_{gf}){\mathcal O}=\sum
\limits_{n=0}^{\infty}\;\int\limits_{\Sigma_\psi}[D\Phi^\ast_A][D\Phi^A]\exp(\frac{\ri}{\hbar} S_{gf}^0)
(S_{gf}^1)^n {\mathcal O}\;,
\end{equation}
where $\mc O$ denotes the considered observable. The splitting of the action into a {\it kinetic} 
 $S_{gf}^0$ and a {\it potential} $S_{gf}^1$ part can be understood as follows. The expansion in  powers of 
$\hbar$ is around the classical solution $X(u)=x$, $A(u)=0$ and one perturb now the fields $X$ by a 
fluctuation field $\xi$, e.g. $ X(u)=x+\xi(u)$. The {\it kinetic part} of the action is then
\beq
S^0_{gf}=\int\limits_{D^2}\;[A_i\wedge(\rd\xi^i+\star\rd\lambda^i)+C_i\rd\star\rd\bar{C}^i]\;.
\end{equation}
The resulting propagator is expressed most easily in the superfield formalism
\beq
\langle\tilde{\xi}^k(w,\zeta)\tilde{A}_j(z,\theta)\rangle=\frac{i\hbar}{2\pi}\delta^k_j\mathcal{D}\phi(z,w)
\end{equation}
with $\phi(z,w)=\frac{1}{2\ri}\ln\frac{(z-w)(z-\bar{w})}{(\bar{z}-\bar{w})(\bar{z}-w)}$ and $z,w$ are 
denoting two points on the disc.

The perturbation series is then obtained by expanding the potential $S^1_{gf}$ in powers of the 
corresponding superfield of the fluctuation $\tilde{\xi}=\xi+\theta^\mu A^\ast_\mu$
\beq S_{gf}^1=\frac{1}{2}\int\limits_{D^2}\int d^2\theta\sum_{k=0}^\infty \frac{1}{k!}\partial_{j_1}\cdots
\partial_{j_k}P^{ij}(X)\tilde{\xi}^{j_1}\cdots\tilde{\xi}^{j_k}\tilde{A}_i\tilde{A_j}\;.
\end{equation}
Now expanding the functions $f(\tilde{X}(1))$ and $g(\tilde{X}(0))$ in powers of $\tilde{\xi}$ yields the 
expansion in {\sc Feynman} diagrams labeled by $\Gamma_n$, where n is the number of vertices. All the 
appearing integrals are absolutely convergent, except those containing tadpole diagrams, a diagram with 
one {\it edge} connecting a vertex to itself, such that one has to take care about renormalization. 
After an usual point-splitting regularization one has to add a resulting counter term which does not 
spoil the classical master equation.

After these considerations one is in a position to calculate the expansion series with the help of the 
{\sc Wick} theorem for {\sc Gauss}ian integrals. The resulting correlation function is then
\beq
(f\star g)(x)=fg+\sum\limits_{n=1}^{\infty}(\frac{\ri\hbar}{2})^n\sum\limits_{\Gamma_n} 
w_{\Gamma_n}{\mathcal D}_{\Gamma_n}(f\otimes g)\;,
\end{equation}
where ${\mathcal D}_{\Gamma_n}$ denotes the resulting bidifferential operator for the diagram $\Gamma_n$ 
coming from the {\sc Poisson} structure on the target manifold and $ w_{\Gamma_n}\approx \int \wedge^n_{j=1}
\rd\phi(u_j,u_{v_1(j)})\wedge\rd\phi(u_j,u_{v_2(j)})$ are the so-called weights which are descended 
from the propagator. For more details on the calculation, in particular the propagator and the boundary 
conditions for the fields, take a look at \cite{CF1}. Note the resulting correlation function is exactly the 
expression for the product in the deformed algebra of functions on an {\sc Poisson} manifold invented by 
{\sc Kontsevich} \cite{K}.
%
%%%%%%%%%%%%%%%%%%%%%%%%%%%%%%%%%%%%%%%%%%%%%%%%%%%%%%%%%%%%%%%%%%%%%%%%%%%%%%%%%%%%%%%%%
%
\subsection{Non-perturbative Aspects}
From now on the world sheet is a closed two-dimensional manifold $M$ with genus $g$.
Important simplifications occur when one writes the action in {\sc Casimir-Darboux} coordinates 
$\{X^I,X^\alpha\}$. The {\sc Batalin-Vilkovisky} action is then
\begin{multline}
S_{BV}= \int\limits_{M}\bigl[A_{ I}\wedge\rd X^{ I}+A_{\alpha}\wedge\rd X^{\alpha}+
\frac{1}{2}P^{\alpha\beta}[X^I]A_{\alpha}\wedge A_{\beta}+\mu C[X^{I}]\\
+A^{\ast I}\wedge\rd C_{ I}+A^{\ast\alpha}\wedge\rd 
C_{\alpha}+X^{\ast}_{\alpha}P^{\beta\alpha}[X^I]C_{\beta}\bigr]\,.
\end{multline}
As one can read of the extended action the gauge freedom of the maps $X^i:M\rightarrow N$ is reduced to the 
freedom of the maps $X^\alpha:M\rightarrow L$ where $L$ is  a symplectic leaf of the {\sc Poisson} 
manifold $N$. 
The gauge transformations reduce to $\delta_\epsilon X^\alpha=(\Omega^{-1})^{\alpha\beta}\epsilon_\beta$ and 
$\delta_\epsilon X^I=0$. Hence, after gauge fixing we need to consider only the homotopy classes 
$[X^\alpha]$. In {\sc Casimir-Darboux} coordinates the gauge fermion can be chosen to be
\beq
\Psi=\int\limits_{M}[\bar{C}^I\chi_I(A_I)+\bar{C}^\alpha\chi_\alpha(X^\alpha)]\;.
\end{equation}
The gauged fixed action in {\sc Casimir-Darboux} coordinates is given by
\begin{multline}
S_{gf}=\int\limits_{M}\bigl[A_{ I}\wedge\rd X^I+A_{\alpha}\wedge\rd X^\alpha
+\frac{1}{2}P^{\alpha\beta}A_{\alpha}\wedge A_{\beta}+\mu C(X^{ I})\\
+\bar{C}^{ J}\frac{\partial\chi_I(A_J)}{\partial A_J}\wedge\rd C_{ I}+
\bar{C}^{ \alpha}\frac{\partial\chi_\beta(X_\alpha)}{\partial X_\alpha}P^{\gamma\beta}C_{\gamma}
-\bar{\pi}^I\chi_I(A_I)-\bar{\pi}^\alpha\chi_\alpha(X^\alpha)\bigr]\;.
\end{multline}
This action can now be used to perform the path integral quantization
\beq
Z=\int\limits_{\Sigma_\Psi}[D\Phi^\ast_A][D\Phi_A]\exp(-\frac{\ri}{\hbar}S_{ gf})\;.
\end{equation}
It is possible to perform all the integrations of the fields \cite{HS1}. Integrating over the ghost and 
antighost fields yields the {\sc Faddeev-Popov} determinants. The integrations over the multipliers yields 
$\delta$-functions which implement the gauge conditions. From now on the integration extends only over the 
degrees of freedom with respect to the gauge-fixing conditions. The integration over $A_\alpha$ is 
{\sc Gauss}ian. Now choose a gauge condition linear in $A_I$, then the {\sc Faddeev-Popov} determinant 
does not depend on $A_I$ anymore and one can integrate over them yielding a $\delta$-function for 
$\rd X^I$. When this $\delta$-function is implemented the fields $X^I$ become independent of the 
coordinates of $M$. Hence the {\sc Casimir} functions are constant and these constant modes $X^I_0$ 
count the symplectic leaves. The gauge fixing of the fields $X^\alpha$ reduces the integral over 
$X^\alpha$ to a sum over the homotopy classes of the maps.
The form of the path integral then becomes   
\begin{multline}
 Z= \int\limits_{\Sigma_\Psi}\rd X^I_0 \sum\limits_{[X^\alpha]}{\rm det}
\left(\frac{\partial\chi_\alpha(X^\alpha)}{\partial X^\gamma}P^{\gamma\beta}(X^I_0)\right)_{\Omega^0(M)}
{\rm det}^{-1/2}\left(P^{\alpha\beta}(X^I_0)\right)_{\Omega^1(M)}\\
\times\exp\left(-\frac{\ri}{\hbar}
\int\limits_M\Omega_{\alpha\beta}\rd X^\alpha\wedge\rd X^\beta\right)\exp\left(-\frac{1}{\hbar}
A_M C(X^I_0)\right)\;,
\end{multline}
where the subscript $\Omega^k(M)$ indicates that the determinant results from an integration over k-forms and 
$A_M$ denotes the surface area of $M$. All the functional integrations have been performed and one has 
arrived at an almost closed expression for the partition function.

Again consider the case in which the manifold $N=\mathbb{R}^3$ and the {\sc Poisson} structure is linear such 
that the coefficients give rise to a {\sc Lie} algebra structure on the dual space $\mc{G}=N^\star$. 
The nontrivial {\sc Casimir} function is again the quadratic one, such that the symplectic leaves are two 
dimensional spheres $\mathbb{S}^2$, characterized, in the {\sc Casimir-Darboux} coordinates, by their 
radius $X^I_0$. 
The symplectic leaves of a linear {\sc Poisson} structure are the co-adjoint orbits of the corresponding 
compact, connected {\sc Lie} group with {\sc Lie} algebra $\mc{G}$. Because the {\sc Lie} algebra has three 
dimensions the {\sc Lie} group is SU(2). By a theorem of {\sc Kirillov} \cite{KL} these orbits can in turn 
be identified with the irreducible unitary representations of G.

Taking into account these simplifications the path integral can be evaluated further. Due to the {\sc Hopf} 
theorem the homotopy classes of the maps $X^\alpha: M\rightarrow \mathbb{S}^2$ are determined by their 
degree $n$. This yields
\beq \int\limits_M \Omega_{ \alpha\beta}\rd X^{\alpha}\wedge\rd X^\beta= n\int\limits_{\mathbb{S}^2}
\Omega_L(X^{ I}_{ 0})\,,
\end{equation}
where $\Omega_L(X^{ I}_{0})$ is the symplectic form on the symplectic leaf L respectively $\mathbb{S}^2$. 
The sum over the degree $n$ yields a periodic $\delta$-function, which says that the symplectic leaves must 
be integral. By the identification of the leaves with the co-adjoint orbits, they must also be integral 
which reduces the number of the co-adjoint orbits to a countable set $\mc{O}(\Omega)$. Choose now the 
{\it unitary gauge} $\chi^\alpha(X)=X^\alpha$ and the two determinants have the same form. Due to the 
{\sc Hodge} decomposition they are characterized by harmonic forms. Now one  
has to count the linear independent forms, which are characterized by the dimension of the corresponding 
homology group, the {\sc Betti} numbers. These yields for the power of the combined determinant the 
{\sc Euler} characteristic $\chi(M)$. Indeed, this is a similar argument used in \cite{BT}. The determinant 
corresponds to the symplectic volume of the symplectic leaf $L_{X^I_0}$. Now one can perform the 
integration over the constant modes $X^I_0$. The partition function of the linear {\sc Poisson}-sigma model 
then takes the form 
\beq
Z=\sum\limits_{\mc{O}(\Omega)}Vol(\Omega)^{\chi(M)}\;\exp(-\frac{1}{\hbar} A_M C(\Omega))\,.
\label{parti}
\end{equation}  

As already pointed out it is possible to identify a {\it linear} {\sc Poisson} manifold with its dual 
space, the {\sc Lie} algebra. This duality leads to the similarity of the partition function of the 
{\it linear} {\sc Poisson}-sigma model (\ref{parti}) and the one of the 
{\sc Yang-Mills} theory. The main tool is the so-called symmetrization map \cite{KL} which maps in this 
case the quadratic {\sc Casimir} $C(\Omega)$ which characterized the co-adjoint orbits into the 
{\sc Casimir} $C(\lambda)$ of the corresponding representation of the {\sc Lie} group. This leads to the 
identification of the integral orbits with the irreducible unitary representations of the {\sc Lie} group. 
The symplectic volume of the co-adjoint orbit equals by a special case of the character formula of {
\sc Kirillov} \cite{KL} the dimension of the corresponding representation. So the partition can be written as
\beq
Z=\sum\limits_{\lambda}\;{\rm d}(\lambda)^{\chi(M)}\exp(-\frac{1}{\hbar}A_{ M}C(\lambda))\;,
\end{equation}
where $\lambda$ denotes the representation while $d(\lambda)$ stands for its dimension. This is exactly 
the partition function for the {\sc Yang-Mills} theory on closed two-dimensional manifolds \cite{BT}.
%
%
%%%%%%%%%%%%%%%%%%%%%%%%%%%%%%%%%%%%%%%%%%%%%%%%%%%%%%%%%%%%%%%%%%%%%%%%%%%%%%%%%%%%%%%%%%
%
\section{Concluding Remarks}
In this review I presented some interesting aspects of the {\sc Poisson}-sigma model discovered in the last 
years. From the mathematical point of view the most interesting fact is the connection with the 
deformation quantization of {\sc Poisson} manifolds \cite{CF1}. On the other hand for physicist the 
important aspect is of course the unified framework for different topological and semi-topological field 
theories \cite{SS} based on the {\sc Poisson} structure on the target manifold. In the special case of a linear
 {\sc Poisson} structure this property could be 
established also on the quantum level by performing the path integral quantization, i.e. it was 
shown that the partition function of the  {\it linear} {\sc Poisson}-sigma model is in some sense the {\it dual} of the one 
of the {\sc Yang-Mills} theory \cite{HS1}.

An interesting step further a whole quantization of the {\sc Poisson}-sigma model would be to calculate the 
partition function on a base manifold with boundaries, in particular on the disc  at least for a linear 
{\sc Poisson} structure \cite{HS2}. The required boundary conditions for the fields must lead to a kind of 
character formula like the one by {\sc Kirillov} \cite{KL} in turn to identify the result with the one for 
the {\sc Yang-Mills} theory where the boundary condition leads to the character of the representation.
The hope  of this considerations is that it would shed some light on the non-perturbative nature of the deformation 
quantization.
\bigskip

%
%
%%%%%%%%%%%%%%%%%%%%%%%%%%%%%%%%%%%%%%%%%%%%%%%%%%%%%%%%%%%%%%%%%%%%%%%%%%%%%%%%%%%%%%%%%%
%
\noindent
{\bf Acknowledgment}
\indent
This work was supported by the {\it Deutsche Forschungsmeinschaft} in connection with the Graduate College for 
Elementary Particle Physics in Dortmund which gave me the possibility to take part at the {\it 3rd International 
Conference on Geometry, Integrability and Quantization} held in Varna, Bulgaria, June 14-23 2001. 
%
%%%%%%%%%%%%%%%%%%%%%%%%%%%%%%%%%%%%%%%%%%%%%%%%%%%%%%%%%%%%%%%%%%%%%%%%%%%%%%%%%%%%%%%%%%
%

\end{document}